# Optical properties of Mn$^{4+}$ ions in GaN:Mn codoped with Mg acceptors

B. Han, R. Y. Korotkov, and B. W. Wessels[a]

*Department of Materials Science and Engineering and Materials Research Center, Northwestern University,*

*Evanston, Illinois 60208*

M. P. Ulmer

*Department of Physics and Astronomy, Northwestern University, Evanston, Illinois 60208*

The optical properties of Mn-Mg codoped epitaxial GaN were studied. Addition of Mg acceptors quenches the weak manganese-related photoluminescence (PL) band at 1.3 eV in GaN:Mn and a series of sharp PL peaks are observed at 1 eV in codoped epilayers. The change in PL spectra indicates that Mg addition stabilizes the Mn$^{4+}$ charge state by decreasing the Fermi level. The 1 eV PL peaks are tentatively attributed to intra center transitions involving Mn$^{4+}$ ions. Spin allowed 3d-shell $^4T_2$-$^4T_1$ transitions and their phonon replicas are involved. The relative intensities of the sharp peaks are strongly dependent on the excitation wavelength, indicating the optically active Mn$^{4+}$ centers involved in the separate peaks are different. The temperature dependence of the PL spectrum suggests the presence of at least three distinct Mn$^{4+}$ complex centers.

Electronic Mail: b-wessels@northwestern.edu

III-V diluted magnetic semiconductors have been extensively studied due to the possibility of tuning their magnetic properties by varying the carrier or magnetic ion concentrations.[1] Recent calculations predict the Curie temperature of *p*-type GaN:Mn can be higher than room temperature.[1] Although observation of high temperature ferromagnetism has been reported, GaN:Mn is always highly resistive in that Mn forms a deep acceptor level at $E_V$+1.4-1.8 eV.[2,3] In order to enhance the carrier-mediated ferromagnetism, codoping GaN:Mn with shallow acceptors is proposed.[1,4] A marked increase of the ferromagnetism has been observed in GaN:Mn by Mg codoping.[4] However, the nature of Mn in codoped GaN is not well established (e.g., the charge state of Mn is unknown). Previously a series of photoluminescence (PL) lines at 1 eV was observed in codoped GaN:[5] assuming the $Mn^{3+/4+}$ level was resonant with the valence band (VB), the 1eV PL peaks were possibly due to the spin-forbidden intra-shell transition $^4T_1$-$^6A_1$ of $Mn^{2+}$ ion.[5] Recent calculations, however, indicated the $Mn^{3+/4+}$ level was above the VB.[6] T. Graf *et al.* subsequently proposed $Mn^{4+}$ is stable instead of $Mn^{2+}$ in Mg codoped GaN:Mn.[3]

In this letter, the origin of the 1 eV PL lines observed in codoped GaN was investigated using selective excitation PL spectroscopy.[7,8] The 1 eV transition series is attributed to spin-allowed intra d-shell transition $^4T_2$-$^4T_1$ of $Mn^{4+}$ ions. Mg codoping has strong effects on the local surroundings of Mn ions as indicated from PL spectra: at least three distinct $Mn^{4+}$ centers with different local crystal field have been observed in codoped GaN:Mn.

Mn-doped and Mn-Mg codoped GaN were grown by metal-organic vapor phase epitaxy as previously described.[2,5,9] PL was measured between 20 and 300K. The excitation source was either the 325 nm line of a He-Cd laser, or the 973.8 nm line of a semiconductor laser.

Figure 1 shows the 20K PL spectra of Mn doped and Mn-Mg codoped GaN, respectively. In the GaN:Mn spectrum a weak band at 1.28 eV was observed, with the full width at half maximum (FWHM) of 290 meV. This peak was previously attributed to the spin-forbidden intra d-shell transition $^4T_1$-$^6A_1$ of $Mn^{2+}$ with a long decay time of 7.8 ms.[9] In the GaN:Mn-Mg spectrum the 1.28 eV band is completely quenched; a series of sharp peaks at 1 eV dominates the spectrum with the FWHM of 3-8 meV. No shift of the peak energies (within a resolution of 2 meV) with temperature is observed up to 200 K, indicating these peaks are weakly coupled to the conduction band or VB and related to intra d-shell transitions.[10,11] The 1 eV PL peaks were previously observed and it was suggested that spin-forbidden intra d-shell transition $^4T_1$-$^6A_1$ of $Mn^{2+}$ was involved.[5] Time resolved PL measurements, however, show these sharp peaks decay exponentially at 20K with the lifetimes of 75±10 μsec (their lifetimes are summarized in Table I). These lifetimes are unexpectedly short for a spin-*forbidden* but typical for a spin-*allowed* intra d-shell transition.[10,11]

According to Tanabe-Sugano diagram, spin allowed d-shell transitions are not observed for a $d^5$ configuration ($Mn^{2+}$) in a tetrahedral crystal field (GaN).[10,11] Consequently, the observed sharp lines must be related to Mn ions with a different charge state. Adding Mg to GaN decreases the Fermi energy, thus the formation of $Mn^{4+}$ is expected in heavily Mg codoped GaN when the Fermi energy is below the $Mn^{3+/4+}$ donor level, which is calculated to be located at 1.1 eV above the VB.[6] The sharp PL peaks in GaN:Mn-Mg are thereby tentatively attributed to spin-allowed intra d-shell transition $^4T_2$-$^4T_1$ of $Mn^{4+}$ ($3d^3$) in a tetrahedral crystal field.[3,10,11] This attribution is consistent with the significant (~two orders of magnitude) PL intensity increase upon Mg codoping (Fig. 1) in that the spin-allowed transition (1 eV peaks in GaN:Mn-Mg) is presumably much stronger than the spin-forbidden transition (1.28 eV peak in GaN:Mn).[10,11]

As shown in the inset of Fig. 1, a series of weaker PL peaks are observed on the low energy side of the codoped film. These peaks are attributed to phonon replicas of the major peaks on the high energy side. In the inset, the major peaks and their phonon replicas are linked by the arrows. The energy positions of major peaks, their phonon replicas, and the corresponding phonon energies are summarized in Table I. From Fig. 1, the phonon energy is determined to be 64±1 meV and attributed to the *lattice* $A_1$(TO) phonon mode of GaN.[12] Although the *pseudolocal* instead of *lattice* phonon modes of Mn ions were observed in GaN:Mn due to the high Mn concentration,[2] *lattice* phonon modes are expected in codoped GaN due to the lower Mn concentration.[4]

Previous work by our group has shown in codoped films the relative intensities of the sharp lines vary with Mn concentration.[5] This variation suggests these peaks are related to different $Mn^{4+}$ centers, in the form of defect complexes, whose relative concentrations depend on Mn concentration. To determine the number and nature of the different $Mn^{4+}$ centers, PL spectra were measured using different excitation photon energies and temperatures.

Figure 2 shows the 20K PL spectra of codoped GaN pumped by (a) below gap and (b) above gap excitations. Although the major sharp peak positions are the same (indicated by the dotted lines), their relative intensities depend strongly on the excitation wavelength: some (e.g. the 1.0318 eV) PL peaks are more efficiently pumped by above gap excitation than other (e.g. the 0.9996 eV) peaks. The selective enhancement of certain transitions indicates these PL transitions are related to different $Mn^{4+}$ centers.[8]

The formation of multiple $Mn^{4+}$ complex centers is further supported by temperature dependent studies of the PL spectra. Figure 4 shows quenching curves of the six major peaks at 1.0579, 1.0478, 1.0318, 1.0130, 0.9996 and 0.9876 eV. The quenching behavior of the latter four can be described assuming the presence of a competing non-radiative process[13]

$$I(T) = I_0 /[1 + A\exp(-E_A / kT)] \qquad (1)$$

where $E_A$ is activation energy of the competing process. As shown in Fig. 3, the 1.0130 and 0.9996 eV peaks quench with the same $E_A = 29\,meV$, indicating both peaks are related to the same $Mn^{4+}$ center. This hypothesis is further supported by the fact that their relative intensities remain invariant when different excitation sources are used (Fig. 2). Similarly the 1.0318 and 0.9876 eV peaks are related to the same $Mn^{4+}$ sites in that (i) both peaks thermally quench with $E_A = 40\,meV$, and (ii) their dependences on excitation wavelength are similar (Fig. 2).

The temperature dependent behavior of the 1.0579 and 1.0478 eV peak intensities, however, cannot be analyzed using Eq. (1): the 1.0579 eV peak intensity increases with temperature up to 80K; temperature dependence of the 1.0478 eV peak is better described assuming the presence of *two* competing non-radiative processes

$$I(T) = I_0 /[1 + A_1 \exp(-E_{A1}/kT) + A_2 \exp(-E_{A2}/kT)] \qquad (2)$$

where $E_{A1}$ and $E_{A2}$, the activation energies of the two competing processes, are determined to be 20 and 10 meV respectively. Based on their quenching behavior, the two peaks are attributed to the same $Mn^{4+}$ center where a splitting of the excited state ($^4T_2$) occurs. The 1.0579 (1.0478) eV peak is attributed to PL emission from the upper (lower) excited state to the ground state. The magnitude of the splitting, determined from the energy separation between the two peaks, is 10.1 meV. With increasing temperature, the population of the upper state increases due to thermal redistribution between the split excited states, resulting in an increase of the 1.0579 eV peak intensity. Meanwhile thermal excitation from the lower to upper excited state accounts for $E_{A2}$ observed in thermal quenching of the 1.0478 eV peak. It should be noted that $E_{A2} = 10\,meV$, which agrees well with the splitting (10.1 meV) of the excited states. Association of the two peaks with the same $Mn^{4+}$ is further confirmed by their similar dependences on excitation wavelength (Fig. 2).

Dependencies of the PL spectra on temperature and excitation wavelength reveal the presence of multiple $Mn^{4+}$ centers in codoped GaN, each characterized by different local environments attributed to the formation of Mn complexes. The complexes could be either Mn-Mg[14] or Mn atomic clusters.[15] In heavily codoped GaN, the formation of Mn-Mg complex is expected in that: i) assuming the concentration $[Mn]=[Mg]=2\times10^{19} cm^{-3}$,[13] the average distance between Mn and Mg atoms is less than 1.5 nm;[14] ii) the formation of a Mn-Mg complex is enhanced by the Coulomb attraction between $Mn_{Ga}^+$ and $Mg_{Ga}^-$.[14,16]

Different Mn-$Mg_j$ (j=1,2…) complexes are formed where j is the number of Mg atoms in the complex. Formation of Mn-$Mg_j$ changes the local crystal field of $Mn^{4+}$ ions, which shifts the $^4T_2$-$^4T_1$ transition energy from that observed in isolated $Mn^{4+}$ centers. A large number of PL peaks are therefore observed when different complexes are involved.

The excitation mechanisms of the 1 eV PL peaks are proposed as follows:

i) <u>980 nm excitation:</u> A charge transfer process is involved following[17]

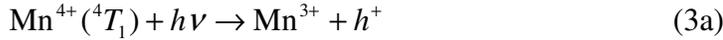
$$Mn^{4+}(^4T_1) + h\nu \rightarrow Mn^{3+} + h^+ \qquad (3a)$$

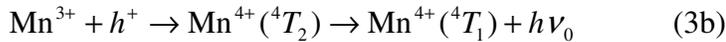
$$Mn^{3+} + h^+ \rightarrow Mn^{4+}(^4T_2) \rightarrow Mn^{4+}(^4T_1) + h\nu_0 \qquad (3b)$$

Electrons are excited from the VB to the $Mn^{4+}(^4T_1)$ state by the 980nm excitation (the $Mn^{3+/4+}$ energy level is 1.1 eV above the VB[6]), forming $Mn^{3+}$ centers (Eq. 3a). Recombination of holes with $Mn^{3+}$ subsequently excites 3d electrons to $Mn^{4+}(^4T_2)$ states, followed by the $^4T_2$-$^4T_1$ radiative transitions at 1 eV (Eq. 3b). A similar excitation process has been observed for intra d-shell transition of $Fe^{3+}$ in GaN.[17]

ii) <u>325 nm excitation:</u> With the above gap excitation, free electrons and holes are created. The electron is trapped at the $Mn^{4+}$ center, subsequently capturing a hole and forming a $Mn^{4+}$ bound exciton. Recombination of this exciton excites the 3d electrons of $Mn^{4+}$ to the $^4T_2$ state through an Auger type process.[7] The observed 1.0 eV peak is a result of the $^4T_2$-$^4T_1$ radiative transition of $Mn^{4+}$. Similar excitation process has been observed for the intra shell transition of $Er^{3+}$ in GaP.[7] It's important to note the 325 nm excitation can be explained by an alternative model involving a Dexter energy transfer.[18] To differentiate between these two models, understanding of the detailed excitation process measured by PL excitation spectroscopy is required. This will be reported separately.

In summary, optical properties of Mn-Mg codoped epitaxial GaN were studied. PL spectroscopy indicates addition of Mg acceptors stabilizes the $Mn^{4+}$ state. A series of sharp PL peaks are observed at 1 eV and attributed to the intra 3d-shell transitions $^4T_2$-$^4T_1$ of the ionized $Mn^{4+}$ ions. The lifetime of these transitions is 75±10 μsec, which is consistent with the spin-allowed transition. Dependence of the PL peak intensity on temperature and excitation energy indicates the peaks are associated with at least three different $Mn^{4+}$ ion complexes, which are tentatively attributed to the formation of Mn-$Mg_j$ centers.

This work is supported by NASA grant NAG5-1147, NSF SPINS program under grant ECS-0224210, and ONR grant No. N00014-01-0012.

[16] Alternatively the multiple $Mn^{4+}$ center complexes can also be attributed to the formation of different Mn atomic clusters (Ref. 15). However, Coulomb interaction presumably favors the formation of Mn-Mg complexes over Mn clusters. In Mn-Mg complexes the attractive Coulomb

interaction between positively charged $Mn_{Ga}^+$ and negatively charged $Mg_{Ga}^-$ ions stabilizes the complex energy.

[17] R. Heitz, P. Maxim, L. Eckey, P. Thurian, A. Hoffmann, I. Broser, K. Pressel, and B.K. Meyer, Phys. Rev. B **55,** 4382 (1997).

[18] In the Dexter model, the optically excited free carriers are first captured by impurity- or defect-related traps (sensitizers). The energy stored in the sensitizers is subsequently transferred by a Dexter non-radiative mechanism to nearby $Mn^{4+}$ centers, inducing a ground to excited state excitation, followed by the 1 eV $^4T_2$-$^4T_1$ PL emission of $Mn^{4+}$ (Ref. 8). Since Mg is the dominant impurity in codoped GaN, the sensitizer is presumably Mg. Similar excitation process has been observed for $Er^{3+}$ in GaN and Mg codoped GaN (Ref. 8).

Table I

Energy positions of the major PL peaks, their lifetimes, their phonon replicas and the corresponding phonon energies in GaN:Mn-Mg.

| Major peak energy (eV) | 1.0579 | 1.0478 | 1.0345 | 1.0318 | 1.0130 | 1.0078 | 0.9996 | 0.9876 |
|---|---|---|---|---|---|---|---|---|
| Lifetime (μsec) | 80 | 84 | 70 | 66 | 67 | 71 | 65 | 71 |
| Phonon replica energy (eV) | 0.9936 | 0.9836 | 0.9695 | 0.9679 | 0.9486 | 0.9444 | 0.9354 | 0.9231 |
| Phonon energy (meV) | 64.3 | 64.2 | 65.0 | 63.9 | 64.4 | 63.4 | 64.2 | 64.5 |

*Figure Captions:*

FIG. 1. Low temperature PL spectra of a Mn doped (RK 342) and a Mn-Mg codoped (RK338) epilayers. In the inset, the major sharp peaks (for codoped GaN) and their phonon replicas are linked by the arrows. The energies of the sharp peaks, their phonon replicas, and the corresponding phonons are summarized in Table I.

FIG. 2. Comparison of the intra d-shell transition of $Mn^{4+}$ in the PL spectra excited by (a) below gap (973.8 nm semiconductor) and (b) above gap (325 nm He-Cd) laser lines.

FIG. 3. Quenching curves of the major sharp peaks. In Fig. 2 these peaks are labeled with the same symbols used for the quenching curves shown in Fig. 3. The activation energies of competing non-radiative processes are determined according to Eq. (1) and Eq. (2). The dashed lines are fitted curves.

Figure 1

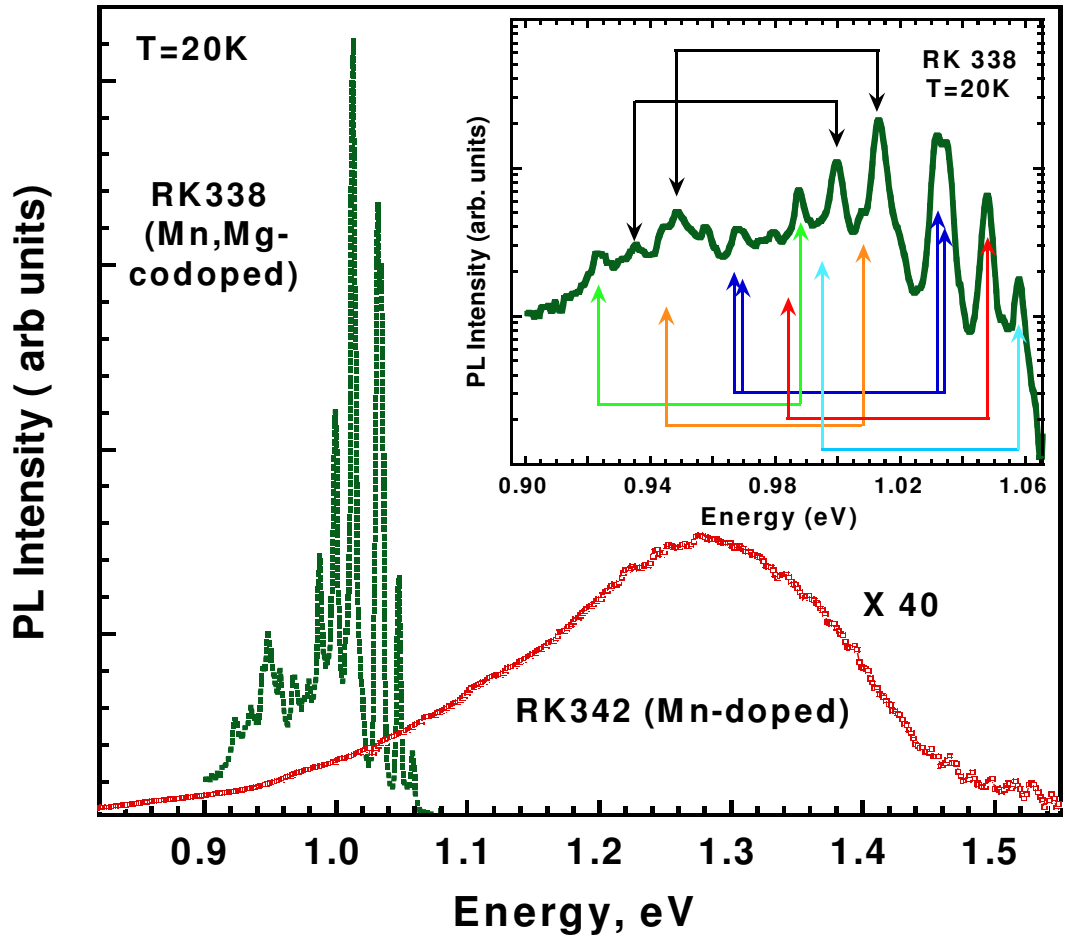

Figure 2

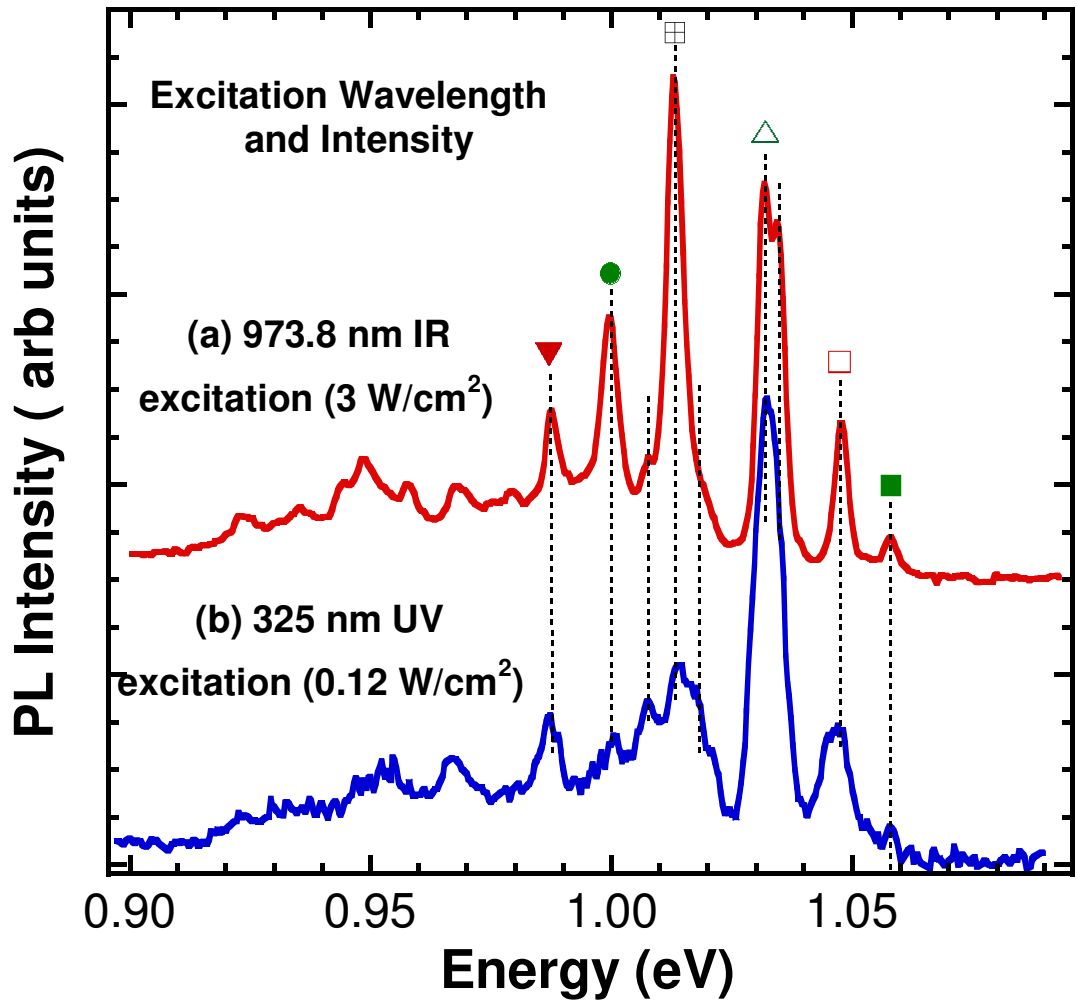

Figure 3

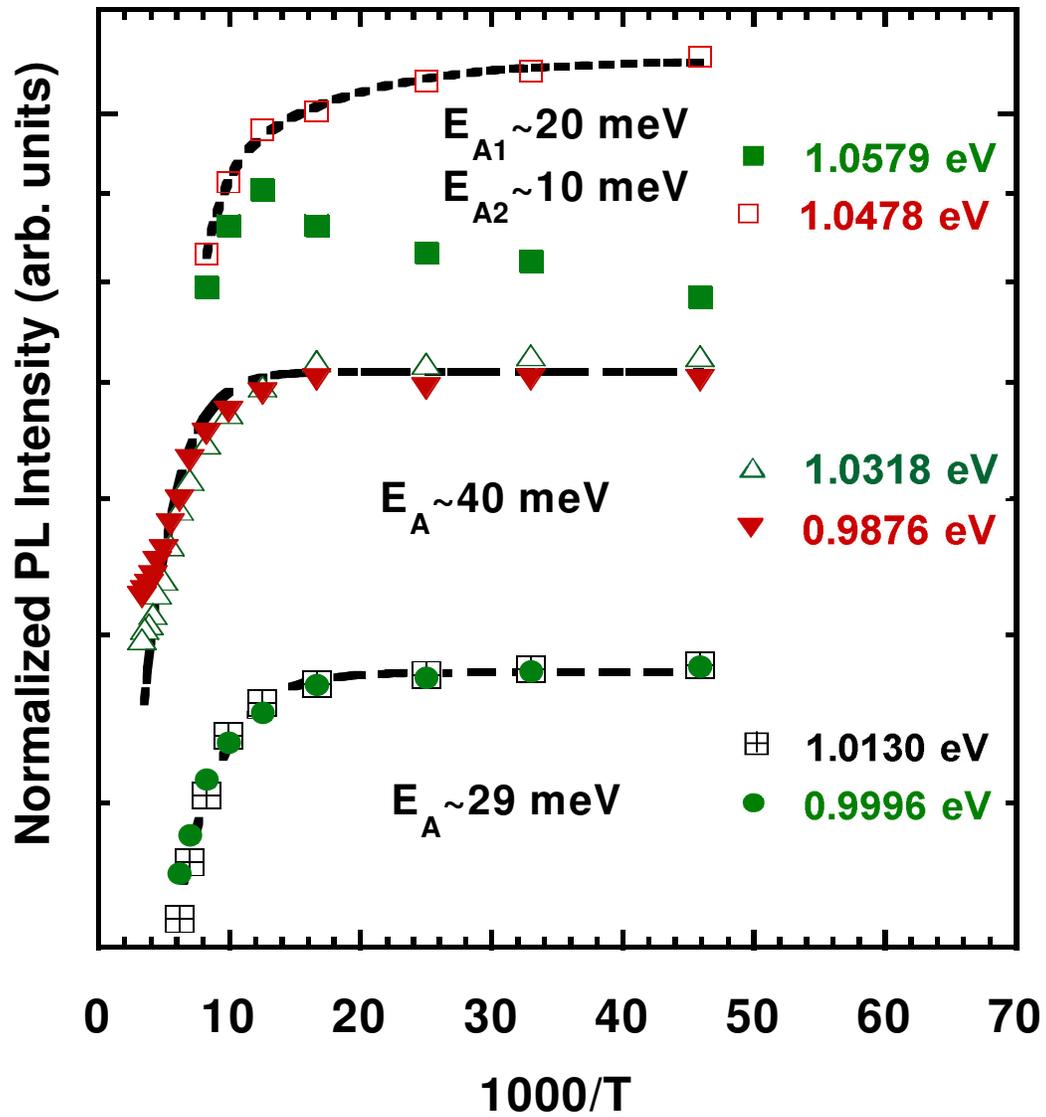